\documentclass[preprint,superscriptaddress,showpacs,preprintnumbers,amsmath,amssymb]{revtex4}
\usepackage{graphicx}

\begin{document}
\title{Strain-dependent damping in nanomechanical resonators from thin  $\mathrm{MoS_2}$ crystals}
\author{E. Kramer}
\affiliation{Kavli Institute of Nanoscience, Delft University of Technology, Lorentzweg 1, 2628 CJ Delft, The Netherlands}
\author{D. J. J. van Dorp}
\affiliation{Kavli Institute of Nanoscience, Delft University of Technology, Lorentzweg 1, 2628 CJ Delft, The Netherlands}
\author{R. van Leeuwen}
\affiliation{Kavli Institute of Nanoscience, Delft University of Technology, Lorentzweg 1, 2628 CJ Delft, The Netherlands}

\author{W. J. Venstra} 
\affiliation{Kavli Institute of Nanoscience, Delft University of Technology, Lorentzweg 1, 2628 CJ Delft, The Netherlands}
\affiliation{Kavli Institute of Nanoscience, Delft University of Technology, Lorentzweg 1, 2628 CJ Delft, The Netherlands; Quantified Air, Lorentzweg 1, 2628 CJ Delft, The Netherlands.}

\date{\today}
\begin{abstract} 
We investigate the effect of mechanical strain on the dynamics of thin $\mathrm{MoS_2}$ nanodrum resonators. Using a piezoelectric crystal, compressive and tensile biaxial strain is induced in initially flat and buckled devices. In the flat device, we observe a remarkable strain-dependence of the resonance line width, while the change in the resonance frequency is relatively small. In the  buckled device, the strain-dependence of the damping is less pronounced, and a clear hysteresis is observed. The experiment suggests that geometric imperfections, such as microscopic wrinkles, could play a role in the strong dissipation observed in nanoresonators fabricated from 2-D materials. 
\end{abstract}
\maketitle
\indent\indent Nanomechanical resonators fabricated from 2-dimensional layered materials, such as graphene and $\mathrm{MoS_2}$, are known to exhibit low quality (Q-) factors at room temperature~\cite{bunch07,chen09,eichler11,barton11,lee13,castellanos13,leeuwen14}. The spectral Q-factor of these devices is orders of magnitude below the values that can be achieved with top-down fabricated devices, such as silicon nitride nanostrings~\cite{verbridge08}. Time-domain measurements on $\mathrm{MoS_2}$ resonators with a thickness down to a single layer, revealed that the low spectral Q-factor is in agreement with the energy relaxation rates~\cite{leeuwen14}, indicating that the line-width is limited by dissipative processes. Although several mechanisms have been proposed  for the high dissipation, such as clamping losses, surface effects, and energy leakage to other vibrational modes~\cite{barton11,croy12,eriksson13,laitinen14,midtvedt14,edblom14}, the dominant mechanism responsible for the excessive dissipation is not identified.\\
\indent\indent It is well known that the Q-factor of top-down fabricated micro-and nanomechanical resonators (MEMS and NEMS resonators) can be increased by introducing tensile strain~\cite{verbridge07,unterreithmeier10, kim10,ning14}. It is explained by considering a complex elastic modulus, $\mathrm{E= E_1+iE_2}$, where the real part, $\mathrm{E_1}$, corresponds to the Hooke\textsc{\char13}s law spring constant, and the imaginary part, $\mathrm{E_2}$, gives rise to dissipation (energy loss). The intrinsic Q-factor can then be written as $\mathrm{Q = E_1/E_2}$. Applying tension increases the real (conservative) part of the elasticity. This results in an increase in the resonance frequency, which is proportional to $\mathrm{\sqrt{E_1}}$, and an increased Q-factor, which is proportional to the resonance frequency by $\mathrm{Q = f_0 / linewidth}$. Previous studies have shown that in MEMS devices the imaginary part of the elasticity can be assumed constant, i.e., independent of strain~\cite{schmid08,unterreithmeier10,schmid11}, and that the tensile strain enhances the Q-factor via the real part of the elasticity, leaving the line-width of the resonance peak virtually unaffected~\cite{unterreithmeier10,yu12}. Since strain engineering is commonly applied to realize MEMS resonators with high Q-factors, it is interesting to investigate the strain-dependence of the Q-factor of mechanical resonators from 2-D materials, which are a 10 to 1000 times thinner, and exhibit very low Q-factors at room temperature.\\
\indent\indent Here we study the strain-dependence of the resonant properties of $\mathrm{MoS_2}$ nanodrum resonators. In contrast to tuning the strain by attracting the drum towards an electrostatic gate electrode~\cite{chen09,chen13}, by applying a pressure difference~\cite{bunch08}, or by chemical modification~\cite{zalalutdinov12,cartamil15}, we use a piezoelectric bender to introduce strain. This enables precise control over the strain, and allows one to study the drum dynamics without exerting out-of-plane forces that could affect the shape of the drum. Both tensile and compressive strain can be introduced. Two devices are considered: one that is initially flat, and one that is initially buckled. In the flat device we observe a weak dependence of the resonance frequency on the strain, but a surprisingly strong strain-dependence of the line-width.
This indicates that the tensile strain enhances the Q-factor via a reduction of the dissipative part of the elasticity, $\mathrm{E_2}$. This is in sharp contrast to top-down fabricated MEMS resonators in which the Q-factor enhances through an increase of real part of the elasticity, $\mathrm{E_1}$. In the buckled device, the changes in the Q-factor are less pronounced. Here we observe hysteresis that could indicate a conformational change of the material, and possibly hints at the underlying process that causes the strain-dependent damping.\\ 
\begin{figure}[hb]
\includegraphics[width=85mm]{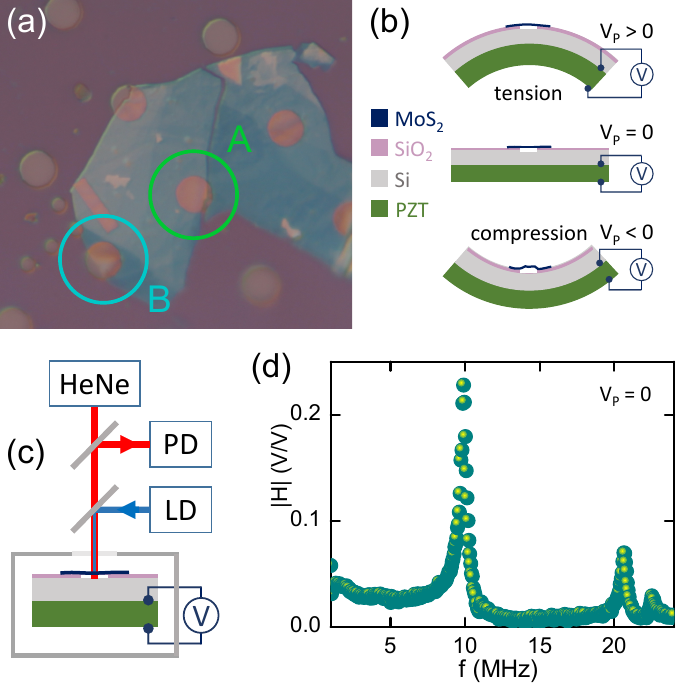}
\caption{(a) Optical photograph of the device. The circles marked A and B indicate the considered devices. The diameter of the holes is $\mathrm{5\,\mu m}$, and the thickness of the flake is $\mathrm{15\,nm}$, which corresponds to ~30 layers. (b) Setup for strain-tuning of mechanical resonators. A thin oxidized Si wafer containing the $\mathrm{MoS_2}$ flakes is glued onto a sheet of piezoelectric material. PZT: Lead Zirconate Titanate piezoelectric material. (c) Interferometric displacement detector. HeNe: Helium-Neon laser; PD: photodiode; LD: laser diode. (d) Frequency response showing the lowest vibrational modes of drum A with $\mathrm{V_P = 0}$. The relation between the resonance frequencies $\mathrm{f_2/f_1\approx 2}$, which indicates the drum behaves mechanically like a plate.}
\end{figure}
\indent\indent To fabricate suspended $\mathrm{MoS_2}$ resonators, we start with a $\mathrm{100\,\mu m}$ thin silicon wafer with a 285 nm thick layer of thermally grown silicon oxide. Thin Si wafers have a low bending rigidity, and this enables the generation of significant mechanical strain. Circular holes are etched in the silicon oxide by conventional electron beam lithography and dry etching. $\mathrm{MoS_2}$ flakes are mechanically exfoliated and deposited onto the substrate using a dry transfer method~\cite{gomez13b}. Figure 1(a) shows the fabricated device; the diameter of the considered drums, marked  A and B, is $\mathrm{5\,\mu m}$, and the thickness is $\mathrm{15\,nm}$, which corresponds to $\mathrm{\approx 30 }$ layers.\\
\indent\indent The wafer containing the drum resonators is fixed onto a commercially available Lead Zirconate Titanate (PZT) piezoelectric sheet, with electrodes on top and bottom. The wafer and the piezoelectric sheet form a bimorph structure, as is shown in Fig.1(b), which bends when an electric field, $\mathrm{V_P}$, is applied across the piezoelectric sheet. Depending on the polarity, compressive or tensile strain is generated in the $\mathrm{MoS_2}$ drum. The motion of the resonator is detected using an optical interferometer, shown schematically in Fig.1(c). The suspended part of the $\mathrm{MoS_2}$ flake forms the moving mirror, while the silicon substrate acts as the reference mirror. When probing the drum with a Helium-Neon laser, the intensity of the reflected optical signal is modulated by the position of the membrane~\cite{azak07,bunch07,lee13,castellanos13,leeuwen14}, and detected using a photo-diode. To measure the frequency response of the drum, it is driven photo-thermally~\cite{bunch07} using a laser diode ($\mathrm{\lambda=405\,nm }$) with an rf-modulated intensity. The measurements are performed at room temperature and at a pressure of $\mathrm{10^{-4}\,mbar}$.\\
\indent\indent Figure 1(d) shows a measured frequency response of drum A, when the voltage applied across the piezo is zero. The fundamental resonance frequency is detected at $\mathrm{f_0=9.9\,MHz}$, with a Q-factor $\mathrm{Q_0=19}$. The second mode is observed at $\mathrm{f_1=20.7\,MHz}$, with a Q-factor $\mathrm{Q_1=43}$. These low Q-factors are typical for mechanical resonators from 2-D materials at room temperature in vacuum~\cite{bunch07,chen09,eichler11,barton11,castellanos13,leeuwen14}. The ratio between the resonance frequencies for the lowest two vibrational modes, $\mathrm{f_1/f_0\approx 2}$, is in agreement with a plate-like resonator. This is as expected, since the thickness of the resonator is beyond the membrane to plate cross-over, which occurs for $\mathrm{MoS_2}$ at approximately five layers~\cite{castellanos13}. In a plate-like resonator the restoring force arises mainly from the bending rigidity, whereas in a membrane it arises from the tension.\\
\begin{figure}[h]
\includegraphics[width=85mm]{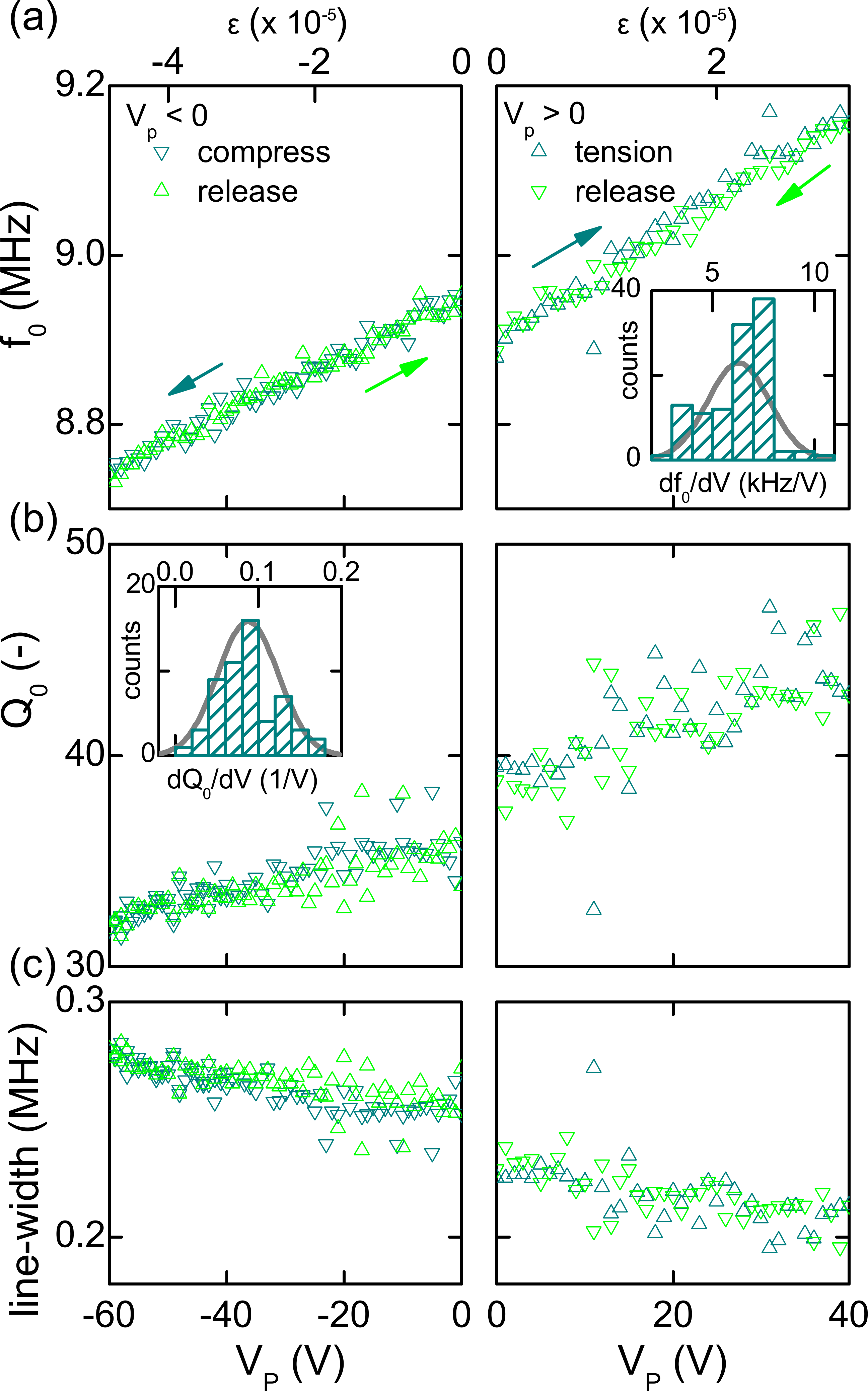}
\caption{Resonance frequency (a) and Q-factor (b) as a function of the voltage applied to the piezo. The panels on the left represent one cycle with compressive strain ($\mathrm{V_P<0}$), the panels on the right one cycle with tensile strain ($\mathrm{V_P>0}$). Multiple cycles were performed, and the (linear) trends of the resulting strain-dependence of the resonance frequency and Q-factor were calculated by least-squares fits to the data, and plotted in the histograms in the insets. Panel (c) shows the corresponding reduction of the line-width of the resonance peak.}
\end{figure}
\indent\indent We now measure the response of drum A while introducing strain by applying a voltage to the piezo. For each voltage the resonance frequency, the Q-factor, and the line-width are obtained from a harmonic oscillator fit. Figure 2 shows the result: the left column represents a compression cycle with $\mathrm{V_P<0}$, and the panel on the right a tensile cycle, with $\mathrm{V_P>0}$. While the strain-dependence of the resonance frequency, shown in panel (a),  is weak, with $\mathrm{\Delta f_0 / f_0\approx 0.02}$, a remarkably strong strain-dependence of the Q-factor is observed, with $\mathrm{\Delta Q_0 / Q_0 \approx 0.25}$ over the same voltage range (panel (b)). The tuning cycle is repeated, and the dependence of $\mathrm{f_0}$ and $\mathrm{Q_0}$ are calculated for each compression and tension cycle, and collected in the histograms shown in the insets~\cite{statistics}. Fitting a Gaussian distribution yields a mean frequency dependence of $\mathrm{\Delta f_0 =6.3\,kHz/V}$ and a Q-factor dependence of $\mathrm{\Delta Q_0 = 0.09\,/V}$. \\
\begin{figure}[h]
\includegraphics[width=85mm]{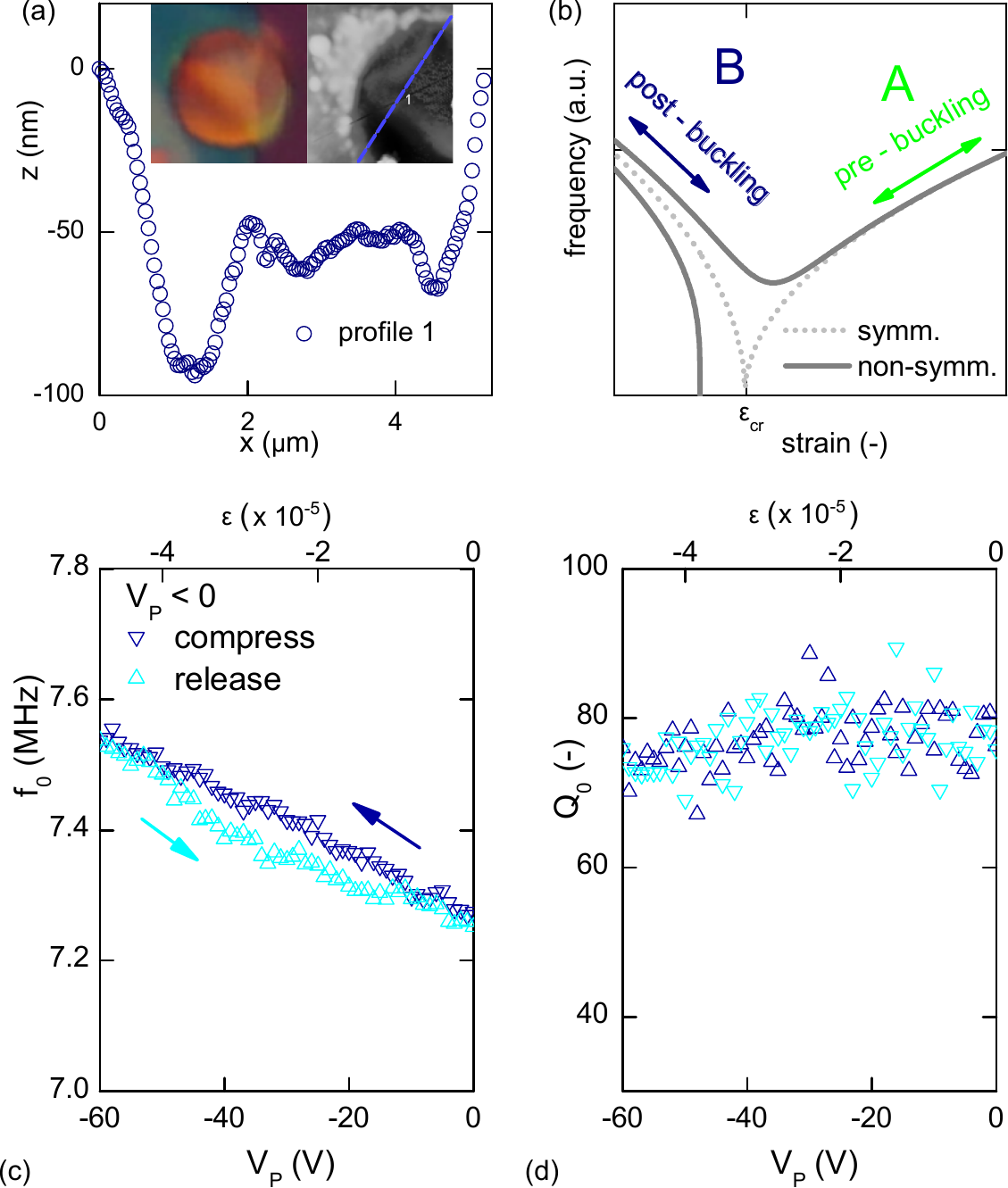}
\caption{(a,inset) optical image (left) and atomic force microscopy topography image (right) of device B. The drum is buckled, as is confirmed by the height profile shown in the main panel. (b) Qualitative behaviour of the resonance frequency in the pre-buckled and post-buckled regimes. Dotted: idealized (symmetric) resonator; solid: realistic (non-symmetric) device. (c,d) Measurements of the tuning of the resonance frequency and Q-factor when varying the compressive strain.}
\end{figure}
\indent\indent To calculate the induced strain as a function of the applied voltage, $\mathrm{V_P}$, we consider a bimorph geometry~\cite{timoshenko25}, with $\mathrm{t_{Si}}$ and $\mathrm{t_{P}}$  the thickness of the silicon and the piezo sheet. The respective Young's moduli are $\mathrm{E_{Si}= 150\,GPa}$ and $\mathrm{E_{P}=62\,GPa}$, and the thicknesses $\mathrm{ t_{Si}= 100\,\mu m}$ and $\mathrm{ t_{P}= 127\,\mu m}$. With the piezoelectric coefficient  $\mathrm{\delta_{31}=-190\times10^{-12}\,V^{-1}}$ and $\mathrm{h=0.534}$ a dimensionless number which represents the ratio's of the Young's moduli and the thicknesses, the strain in the $\mathrm{MoS_2}$ is calculated as $\mathrm{\epsilon=\delta_{31}h/ t_{P} \times V_{P}=8.0\times10^{-7}\,V^{-1}}$.  The calculated strain is plotted on a secondary x-axis in Fig.2. When the compressive strain exceeds a critical limit, the plate buckles. For a circular plate, the critical strain is calculated as $\mathrm{\epsilon_{cr}=\sigma_{cr}/E_{MoS_2}=\frac{K}{1-\nu^2}(\frac{t}{r})^2}$~\cite{roark02}. Here, $\mathrm{E_{MoS_2}}$ is the real part of the Young's modulus, $\mathrm{\nu=0.25}$ is Poisson's ratio, and t and r are the plate thickness and radius. K is a constant that depends on the boundary condition, with $\mathrm{K=1.22}$ for a clamped plate. Although in the present experiments the critical strain should occur at  $\mathrm{\epsilon_{cr}=4.7\times 10^{-5}}$, which corresponds to  $\mathrm{V_{P}=-59\,V}$, no buckling is observed~\cite{vp150v}.\\
\indent\indent We now turn our attention to device B, which is shown in detail in the inset of Fig.3(a). Clearly, a part of the drum is bulged: the bright color in the center indicates a buckle, which is the result of residual compressive strain which is introduced during fabrication. This device allows us to  investigate the strain-dependent behaviour in the post-buckled regime. The topographic AFM image in the second inset, confirms the presence of a buckle: a line-cut across the buckle (main figure) reveals a height of several tens of nanometers, with multiple smaller corrugations and wrinkles superimposed.\\
\indent\indent The presence of wrinkles  make and analytical treatment of the resonance frequencies difficult,  and instead, we present here only qualitatively the strain-dependent resonance frequency of a clamped-clamped plate. Figure 3(b) represents two cases: an idealized symmetric system (grey dots), and a system that more closely resembles the non-symmetric $\mathrm{MoS_2}$ nanodrum (blue solid line). While in the pre-buckled regime the resonance frequency increases with the tensile strain, in the post-buckled regime it is expected to decrease with the tension. These situations apply to the devices considered: device A is in the pre-buckled regime and tunes according to the green arrow, and device B is post-buckled and is expected to tune along the blue arrow.\\
\indent\indent Figure 3(c,d) show the measured tuning behaviour of drum B. Indeed, the frequency dependence is opposite to the one observed for drum A, as the resonance frequency decreases with the tensile strain. Compared to the flat device, the strain-dependence of the Q-factor is less pronounced. The low strain-dependence of the Q-factor could be explained by a relaxation of the compressive stress in the post-buckling regime, where elastic energy is converted from compression to bending. Interestingly, the frequency tuning curve shows a clear hysteresis: when the plate is compressed, the frequency vs. strain response follows a different path than when the strain is released. This cannot be due to hysteresis in the piezo stack, since in the measurements on device A in Fig. 2 the forward and backward tuning curves coincide. The observed effect is attributed to a change in the mechanical properties of the flake.  Hysteresis in the post-buckled regime could indicate a conformational change, possibly of one of the wrinkles. Similar hysteretic effects could occur at a smaller dimensional scale, and give rise to energy dissipation, causing the excessive damping of resonant motion.\\
\indent\indent Another explanation for the strong strain-dependent damping could be the inevitable presence of (static) microscopic corrugations and wrinkles. Theoretical investigations have shown that microscopic geometric artefacts act as long-wavelength elastic scatterers~\cite{helgee14}, carrying away energy from the flexural modes. The wrinkles are not present in top-down silicon-based devices, which are inherently flat due to their fabrication process. Applying tensile strain to the 2-D resonator \textsc{\char13}irons-out\textsc{\char13} the static wrinkles, which reduces the number of scatterers and results in a lower dissipation (i.e., a reduction of $\mathrm{E_2}$), while the resonance frequency ($\mathrm{E_1}$) is affected only weakly. In addition to the static wrinkles, the 2-D material resonators exhibit dynamic wrinkles due to the thermal fluctuations. Applying strain increases the spring constant ($\mathrm{E_1}$), which reduces the mean squared amplitude of these fluctuation-induced dynamic wrinkles. Dynamic wrinkles are far less pronounced in top-down fabricated devices, which are typically thicker by one or two orders of magnitude and therefore have a much higher spring constant. This results in thermal fluctuations with relatively low amplitudes.\\
\indent\indent Besides tuning the damping in mechanical resonators, there are other interesting applications for controlled strain tuning in 2-D materials. In these materials, which can be excessively strained due to the lack of defects~\cite{perez14}, the mechanical strain changes the band structure. The qualitative changes in the electronic and optical properties~\cite{duerloo13,wu14,amorim15} can enable applications such as piezo-electric energy harvesters~\cite{lopez14, wu14} and pressure, motion, and mass sensors~\cite{yan14}. While biaxial strain can be adjusted by varying the temperature by deploying the thermal expansion mismatch~\cite{bao09}, the controlled application of strain described here, can be used to study the strain-dependent properties of 2-D materials in great detail.\\ 
\indent\indent In conclusion, we studied experimentally the strain-dependence of the Q-factor in thin $\mathrm{MoS_2}$ drum resonators in the pre- and post-buckling regime. The experiments indicate that, as in MEMS and NEMS resonators, the Q-factor increases with the applied tensile strain. However, in the $\mathrm{MoS_2}$ resonators, the increase in Q is manifests as a reduction of the line-width, which indicates a decrease in dissipative part of the spring constant. This is in contrast to top-down fabricated MEMS resonators, where Q increases with strain due to an increase in the conservative part of the spring constant, which has only a small effect on the resonance line-width. This result sheds light on the very low Q-factors observed in recent experiments with 2-D mechanical resonators, and suggests that microscopic wrinkles and corrugations, which are ironed-out by applying tensile strain, could play a role in the observed low Q's. In the post-buckled device, hysteresis is observed, and the Q-factor depends less on the strain. The experiment shows that strain engineering is a viable tool to reduce the damping in nanodrum resonators made from 2-dimensional materials.\\
\indent\indent We thank Andres Castellanos-Gomez for assistance with the device fabrication, and Fredrik Creemer and Herre van der Zant for discussions. This work was financially enabled by NanoNextNL, a micro and nanotechnology consortium of the Netherlands and 130 partners, and the European Union's Seventh Framework Programme (FP7) under Grant Agreement $\mathrm{n{\circ}~318287}$, project LANDAUER.\\

\end{document}